\documentclass{emulateapj}
\usepackage{hyperref}

\slugcomment{}

\shorttitle{Red nuggets at intermediate redshift}
\shortauthors{Damjanov et al.}

\begin{document}

\title{Discovery of Nine Intermediate Redshift Compact Quiescent Galaxies in the Sloan Digital Sky Survey}

\author{Ivana Damjanov\altaffilmark{1}, Igor Chilingarian\altaffilmark{2,3}, Ho Seong Hwang\altaffilmark{2}, and Margaret J. Geller\altaffilmark{2}}
\altaffiltext{1}{Harvard-Smithsonian Center for Astrophysics, 60 Garden St., Cambridge, MA 02138}
\altaffiltext{2}{Smithsonian Astrophysical Observatory, 60 Garden St., Cambridge, MA 02138}
\altaffiltext{3}{Sternberg Astronomical Institute, Moscow State University, 13 Universitetsky prospect, 119992 Moscow Russia}

\begin{abstract}
We identify nine galaxies with dynamical masses of M$_{dyn}\gtrsim10^{10}$~M$_\odot$  as photometric point sources, but with redshifts between $z=0.2$ and $z=0.6$,  in the Sloan Digital Sky Survey (SDSS) spectro-photometric database.
All nine galaxies  have archival {\it Hubble Space Telescope} (HST) images. Surface brightness profile fitting confirms that all  nine galaxies are extremely compact ($0.4<$R$_{e,c}<6.6$~kpc with the median R$_{e,c}=0.74$~kpc) for their velocity dispersion ($110<\sigma<340$~km~s$^{-1}$; median $\sigma=178$~km~s$^{-1}$). From the SDSS spectra,  three systems are dominated by very young stars; the other six are older than $\sim1$~Gyr (two  are E+A galaxies).  The three young galaxies have disturbed morphologies and  the older systems have smooth profiles consistent with  a single  S\'ersic function. All nine lie below the $ z \sim 0$ velocity dispersion-half-light radius relation. The most massive system  - SDSSJ123657.44+631115.4 -  lies right within the locus for massive compact $z>1$ galaxies and the other eight objects follow the high-redshift dynamical size-mass relation.      
\end{abstract}

\keywords{galaxies: evolution --- galaxies: fundamental parameters --- galaxies: stellar content --- galaxies: structure}

\section{Introduction}

The observed strong size evolution of massive quiescent galaxies is a fascinating challenge to our understanding of galaxy formation and evolution \citep[e.g., ][]{Khochfar2006, Fan2008, Nipoti2009, Naab2009, Hopkins2010, Oser2010, Ragone-Figueroa2011, Shankar2012}. \citet{Daddi2005} first discovered extremely compact passively evolving systems with half-light radii R$_e<1$~kpc at  redshift $z >1.4$. Further HST observations show that at a fixed stellar mass, galaxies at zero redshift are generally a factor of $3-5$ larger than their high redshift counterparts  \citep[e.g., ][]{Trujillo2007,Toft2007, Zirm2007,Buitrago2008, Cimatti2008, VanDokkum2008, Bezanson2009, Damjanov2009, Damjanov2011,Carrasco2010, Strazzullo2010, Saracco2010, Cassata2011, Szomoru2012, Bruce2012}. 

Velocity dispersions measured for small samples of quiescent systems at high redshift confirm that their dynamical masses agree well with the stellar masses derived from SED-fitting \citep[see ] [ and the references therein]{VandeSande2013}. There are thus two coexisting populations  of massive quiescent systems at $z\gtrsim1$: 1) very dense compact systems and 2) systems with sizes comparable with typical $z\sim0$ quiescent galaxies \citep{Mancini2010, Saracco2010, Cassata2011, Newman2012, Onodera2012}. 

Compact massive systems seem to disappear by $z \sim 0$, but there are conflicting observations. \citet{Trujillo2009} and \citet{Taylor2010} use the SDSS to suggest that the number density of compact massive systems at $z<0.2$ is more than three orders of magnitude  below the comoving density at $z\sim2$.  In contrast, from ground-based imaging combined with spectroscopy, \citet{Valentinuzzi2010a} find a  significant fraction of compact massive galaxies in the WINGS cluster sample at $z\sim0.05$; they derive a lower limit on the number density of $n\sim1.3\times10^{-5}$~Mpc$^{-3}$, comparable with the number density of the high-redshift analogs. A similar study of the field population at $0.03\leqslant z\leqslant0.11$ \citep{Poggianti2013} suggests that compact dense galaxies  exist in this redshift range but their fraction is three times smaller than in the WINGS cluster environment sample.  

There are few observational tests of the existence of compact galaxies at redshifts  $0.1 \lesssim z \lesssim 1$ \citep{Saglia2010, Carollo2013}. \citet{Valentinuzzi2010b} identified compact systems with stellar masses M$_{\ast}>4 \times 10^{10}$M$_\odot$ among spectroscopically confirmed members of rich galaxy clusters with $0.5<z<0.8$.  Here we carry out an environment independent search for compact objects in the redshift range $0.2 < z < 0.6$ by combining  the photometric and spectroscopic SDSS databases with high-resolution images in the Mikulski Archive for Space Telescopes (MAST). Discovery of dense galaxies in this redshift range is important because larger samples with
well-defined selection criteria potentially constrain  models of galaxy evolution.  

We adopt a $\Omega_{\Lambda}=0.73$, $\Omega_{M} = 0.27$, and $H_0 = 70$~km~s$^{-1}$~Mpc$^{-1}$ cosmology, and quote magnitudes in the AB system.

\section{Identifying Compact Galaxy Candidates }\label{sec:data}

We use the SDSS (Release 7; SDSS DR 7) to initiate the search for candidate compact quiescent galaxies in the redshift range $0.2 < z < 0.6$, where the SDSS main sample combined with the BOSS Survey contain spectra for large numbers of objects \citep{Ahn2013}. To identify compact systems we search for objects identified photometrically as stars (in the {\it PhotoObj} view) but with a redshift in our target range (from the {\it SpecObj} view). Thus we obtain a list of object with sizes less than the SDSS PSF ($\sim1\farcs5$). We check that the photometric and spectroscopic objects actually have the same center and that the objects are visually compact in the SDSS images. Additionally we eliminate objects classified spectroscopically as quasars. 

To restrict the list to quiescent galaxies, we require  that the equivalent width of the $[$OII$]\lambda\lambda3726,3729$ emission line doublet is EW$[$O II$]<5$~\AA. We check visually that each spectrum has a clear  4000~\AA~break along with several absorption features (e.g. Balmer series, Ca H+K and G-band). The final list of SDSS DR7 compact system candidates at $0.2<z<0.6$ includes 635 galaxies. 

The SDSS photometric dataset provides only an upper limit to the angular size of these systems, corresponding to a physical radius between 2.5~kpc (at $z=0.2$) and 5~kpc (at $z=0.6$). To obtain direct size measurement we searched the HST archive. Nine of our 635 candidates have HST images.
 
Table~\ref{tab2} lists the camera and the filter for each HST observational program together with the corresponding pixel scale. The exquisite HST resolution of $\sim 0\farcs15$ allows analysis of the structure of these systems on a spatial scale of a few hundred pc ( $\lesssim 500$~pc at $z=0.6$). 

\section{Spectroscopy and Imaging}

\begin{figure*}[h!]
\begin{centering}
\includegraphics[scale=0.35]{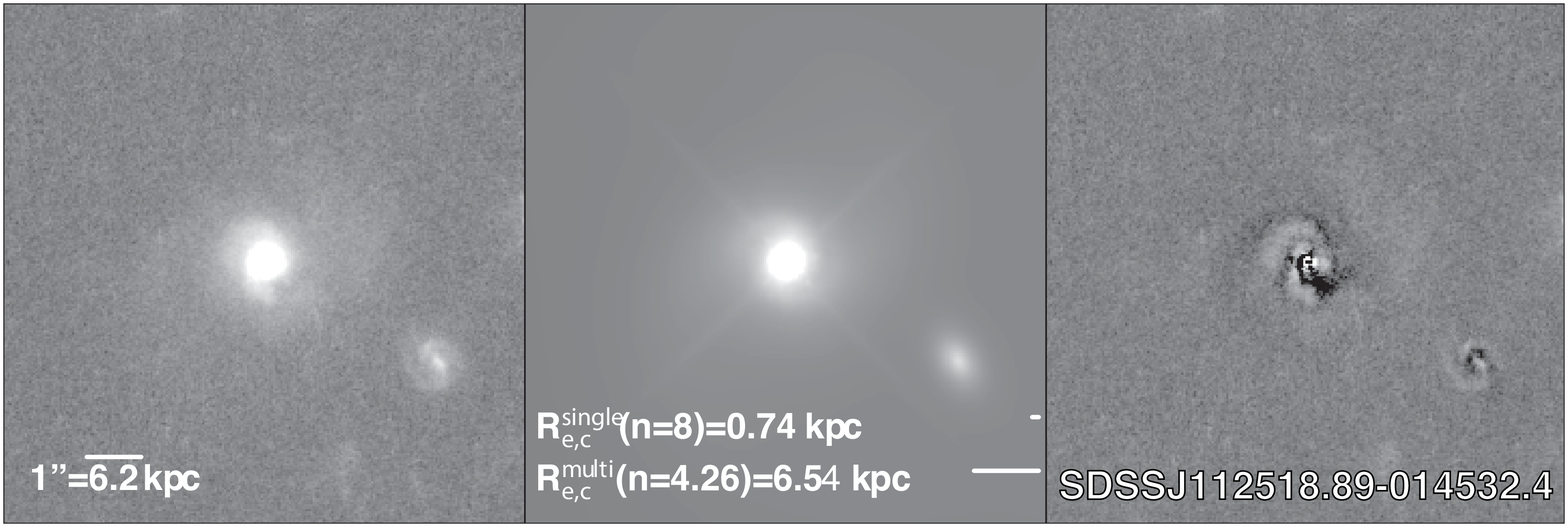}
\hspace*{-.25in}
\includegraphics[scale=0.45]{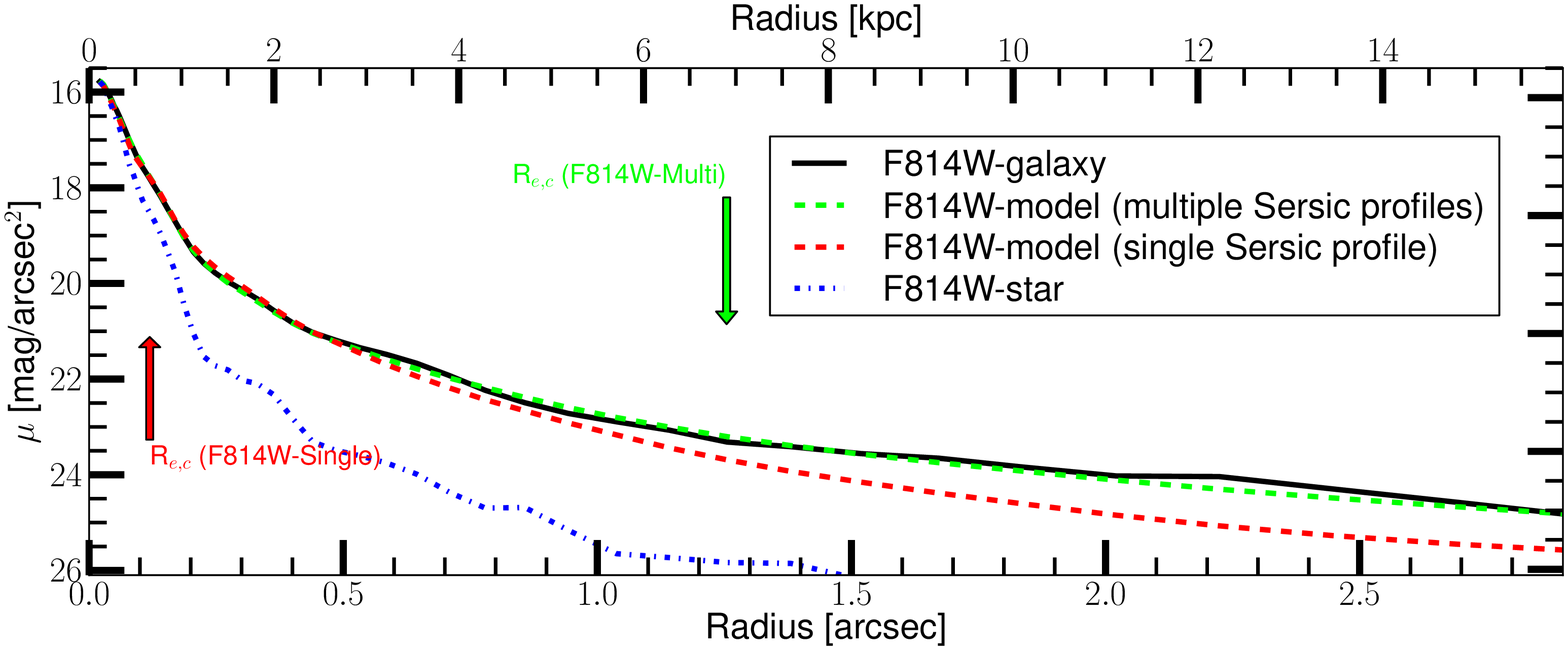}
\hspace*{-.25in}
\includegraphics[scale=0.75]{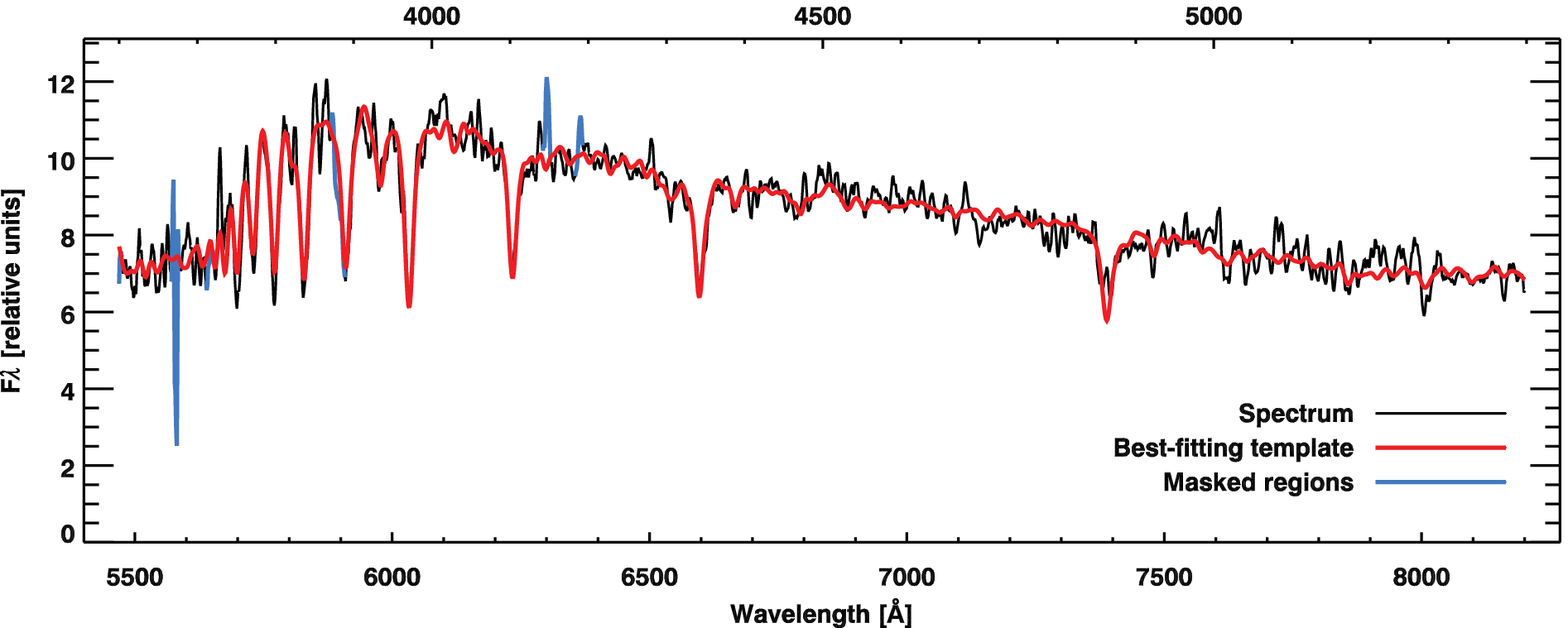}

\caption{\footnotesize  A compact intermediate-redshift galaxy dominated by a young stellar population. We show the HST/WFC3 F814W image (top left), the 2D fit (top center), and the residual (top right). The white bars show relevant scales. The central panel shows the 1D observed surface profile (black), the best single (red) and double (green) S\'ersic fits, and corresponding half-light radii. We also show the PSF (blue). The bottom panel shows the smoothed SDSS spectrum (black), the best-fit SSP model (red), and the regions excluded from the fit (blue). The bottom axis of this panel shows observed wavelength and the top axis gives galaxy rest-frame wavelength. Note the strong Balmer absorption and absence of prominent emission lines.\label{fig1}}
\end{centering}
\end{figure*}

\begin{figure*}[h!]
\begin{centering}
\includegraphics[scale=0.35]{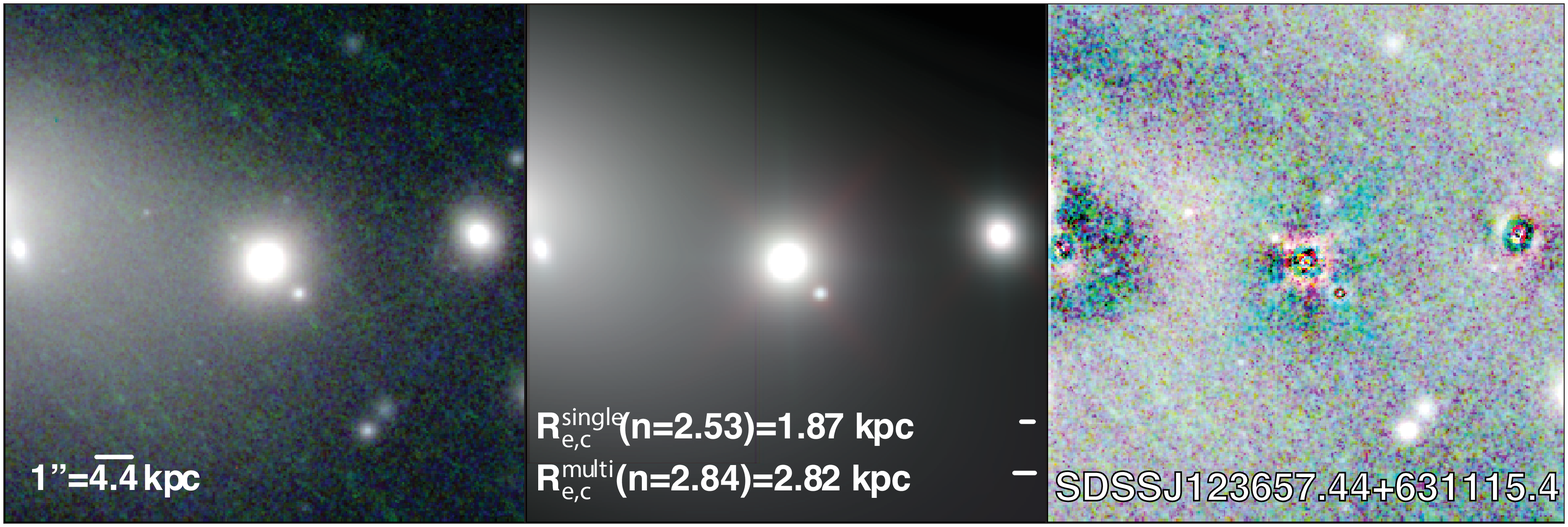}
\hspace*{-.25in}
\includegraphics[scale=0.45]{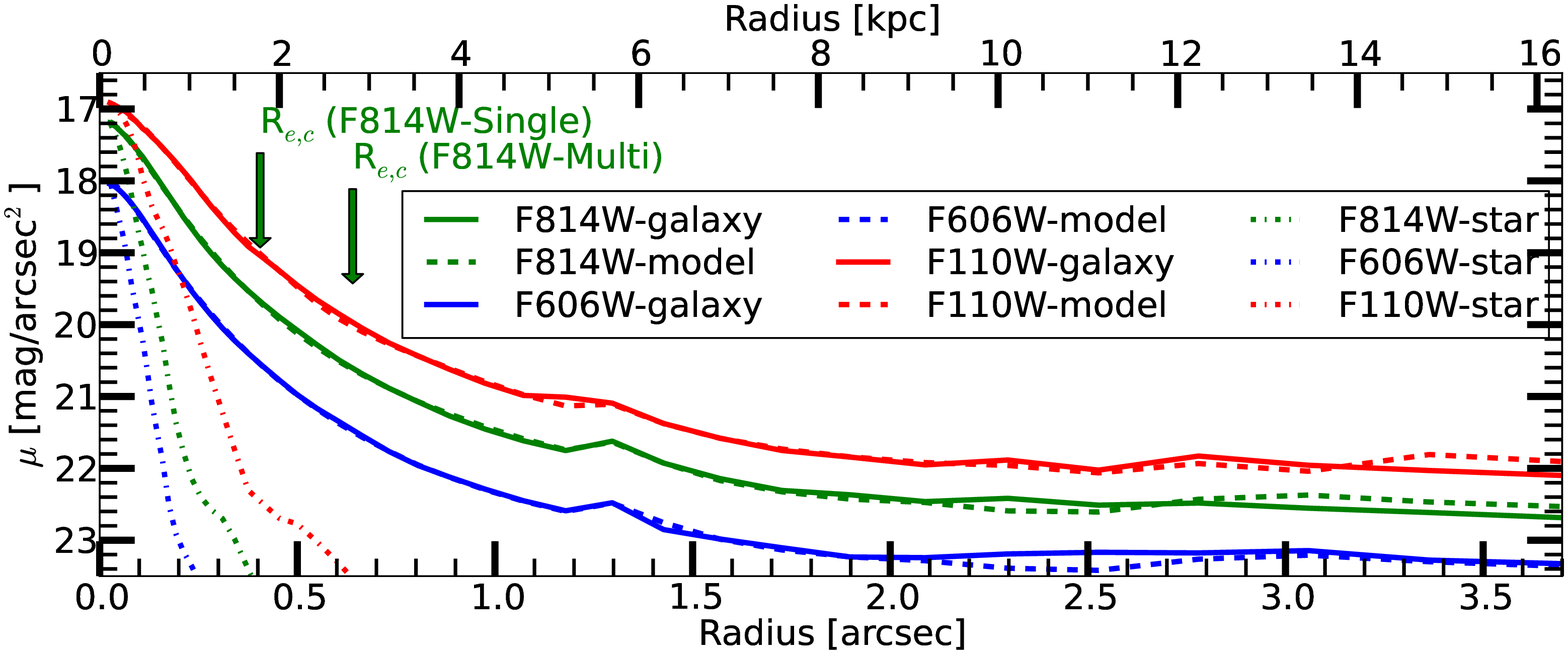}
\hspace*{-.25in}
\includegraphics[scale=0.75]{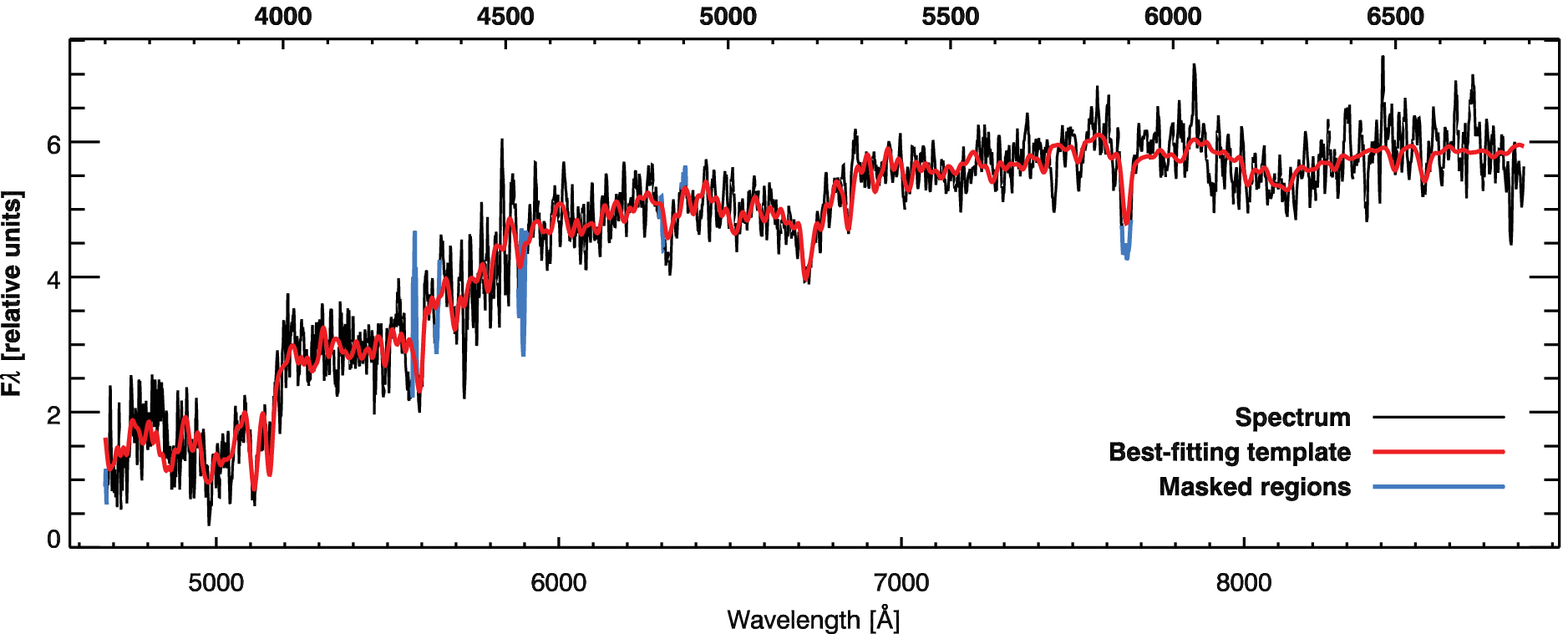}

\caption{\footnotesize A compact massive intermediate-redshift galaxy dominated by an old stellar population. The RGB image (top left) is composed of HST/WFC3 F110W (red), HST/ACS F814W (blue), and HST/ACS F606W (green) light profiles. We also show the RGB image of the best-fit 2D multi-S\'ersic profile models in these three filters (middle), and the three-color residual between the observed surface brightness profiles and the models (right). White bars show relevant scales. The central panel shows  1D radial surface brightness profiles  (solid lines), the best-fit composite models (dashed lines), and the PSF (dashed-dotted lines). Green arrows denote the half-light radii. The bottom panel shows the smoothed SDSS spectrum (black), the best-fit SSP model (red), and the regions excluded from the fit (blue). The bottom axis  of this panel shows observed wavelength and the top axis gives galaxy rest-frame wavelength. \label{fig2}}
\end{centering}
\end{figure*}

\begin{figure*}[h!]
\begin{centering}
\includegraphics[scale=0.5]{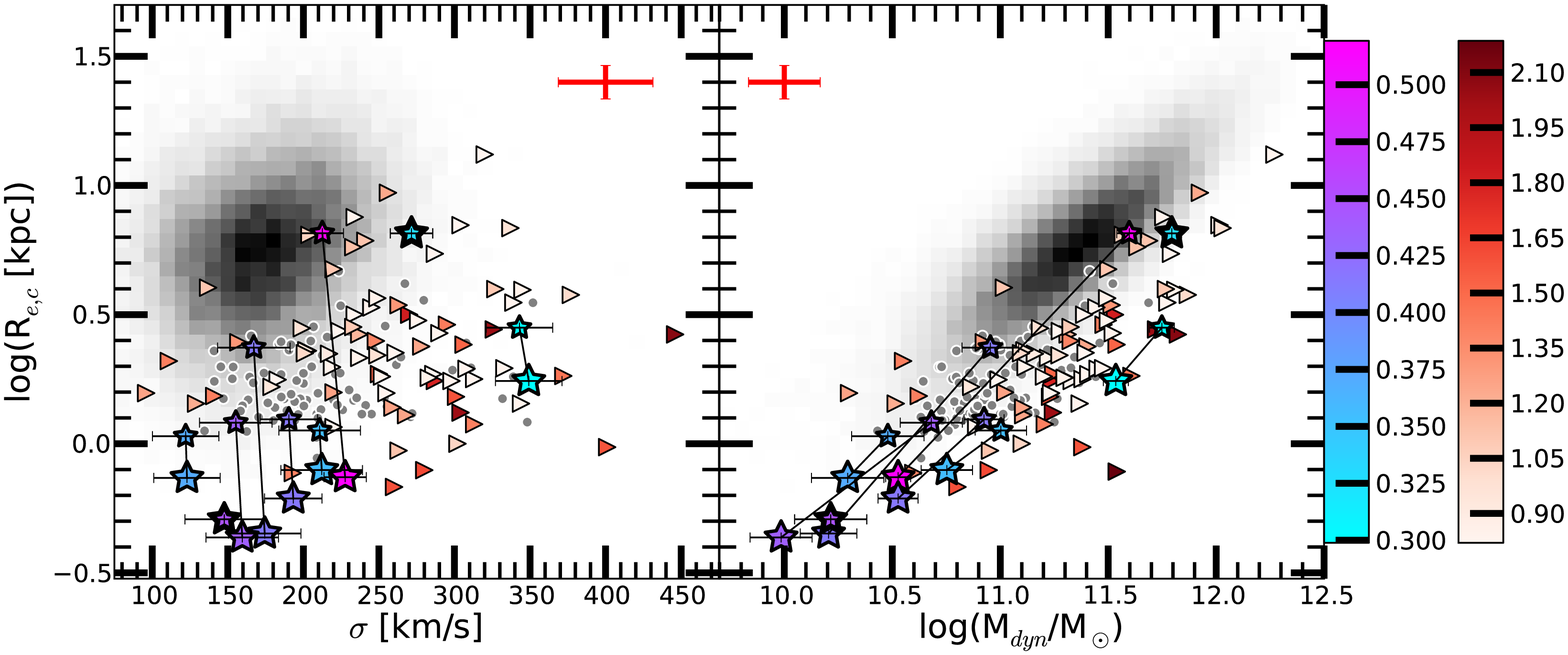}
\caption{\footnotesize The size-velocity dispersion relation {(\it left)} and the size-dynamical mass relation {(\it right)} for quiescent galaxies in three redshift ranges: a) $z\lesssim0.3$ (grey 2D histogram), b) $0.2<z<0.6$ (this study, stars), and c) $0.8<z<2.2$ (triangles). The symbol color indicates galaxy redshifts according to the color bars. Two points connected by a solid line represent each target in our sample: the smaller circularized half-light radii R$_{e,c}$ (larger star) corresponds to the single-S\'ersic profile  and the larger  R$_{e,c}$ (smaller star) denotes the upper limit from the largest R$_{e,c}$ in the multi-S\'ersic profile. The red bar shows the average error for the high-redshift sample. The grey points denote the $z\sim0$ compact sample \citep{Trujillo2009,Taylor2010}.\label{fig3}}
\end{centering}
\end{figure*}

\subsection{The SDSS Spectra} \label{sec:spectralfit}

We reanalyze the SDSS spectrum for each of our nine objects to obtain 
radial velocity, velocity dispersion, mean age and metallicity.
We first fit SDSS DR7 spectra of galaxy candidates against a grid of {\sc
pegase.hr} \citep{LeBorgne+04} simple stellar population (SSP) models based
on the MILES stellar library \citep{Sanchez-Blazquez+06} using the {\sc
nbursts} pixel space fitting technique \citep{CSAP07,CPSK07}.  For every
spectrum we first convolve the SSP model grid covering a wide range of ages
and metallicities with the instrumental response of the SDSS spectrograph.

Next our minimization procedure convolves the SSP models again with a Gaussian line-of-sight velocity distribution,
and multiplies the models by a smooth low-order continuum polynomial aimed at absorbing flux calibration errors in both
models and the data. We choose the best-fitting SSP (or a linear combination of two) by interpolating a grid in age and metallicity.

We repeat the fitting procedure to the entire spectrum and to a spectrum with the regions of known emission lines blocked out.
We analyze  the full available spectral range (from  the rest-frame 3640\AA\ to 6800\AA\ or the red end of the spectrum if it occurs at shorter rest-frame wavelength).  If the reduced
$\chi^2$ value is significantly lower for the emission-line clipped case, the galaxy
has significant emission lines in its spectrum.

In the sample of candidate galaxies, we identify two populations (Table~\ref{tab1}).
Six galaxies resemble classical elliptical galaxies with  ages
$t\gtrsim 1$~Gyr, high metallicities
(from slightly sub-solar to super-solar), and absence of ongoing star
formation. Two of these objects are E+As.

Three blue galaxies have young stellar populations ($t < 50$~Myr), high-velocity outflows,  and sometimes residual star formation
\citep{Diamond-Stanic2012}.  In one case (SDSSJ112518.89-014532.4, Table~\ref{tab1} and Figure~\ref{fig1})  we identify an
``extreme post-starburst galaxy'' which we apparently
observe immediately after the cessation of a strong and short star formation
episode: this object has no prominent  emission lines.  However, the mean
stellar age is $\sim$ 30~Myr, close to the lowest limit covered by
our SSP models. This object is considerably younger than typical E+A galaxies 
with luminosity weighted ages of 500--800~Myr \citep{CdRB09,Du+10}.  

\subsection{HST Imaging Structural Analysis}\label{sec:galfit}

We process dithered images of the nine galaxies in Table~\ref{tab1} using AstroDrizzle\footnote{\url{http://drizzlepac.stsci.edu}}. This  image processing step combines individual exposures and rejects spurious pixels (cosmic rays and hot pixels) without changing the native pixel scale. In the special case of SDSSJ123657.44+631115.4, we process the WFC3 IR image to produce a science mosaic with a pixel scale corresponding to the scale of the ACS images.

We characterize the  HST surface brightness profiles of our candidates by fitting a set of 2D $R^{{1}\over{n}}$ S\'ersic profiles \citep[GALFIT; ][]{Peng2010}. Before fitting, we convolve the models  with the Tiny Tim PSF \citep{Krist2011}. For many of our candidates there are not enough suitable stars to construct a high signal-to-noise PSF. For targets where a number of stars are available in the image, using the Tiny Tim PSF and the stellar PSF produce very similar results. For consistency, we use the Tiny Tim PSF for all images.

To fit the surface brightness profiles,  we start with a single S\'ersic profile and simultaneously fit all objects and the sky background in the $\sim10\arcsec\times10\arcsec$ FoV around our target. With each fitting iteration we  enhance the complexity of the model by adding one more S\'ersic profile. We repeat fitting  until the sky background estimates reach a plateau. \citet{Huang2013} show that this procedure is reliable for large enough regions. If the  residual image does not show any prominent structure, we adopt the  multiple S\'ersic profile as the best fit. With the sky background fixed, we then fit a single S\'ersic profile  to extract structural parameters for direct comparison with the high-$z$ compact passive systems. The resolution at high redshift limits the fit to a single S\'ersic profile. Table \ref{tab2} lists the parameters of the best-fit multi-component and single-component  models.    

Figures~\ref{fig1}~and~\ref{fig2} show the HST images, the best-fit 2D GALFIT models and residuals, the 1D observed and modeled profiles, and the SDSS spectra and fits for a representative young and old object, respectively. We have applied K plus evolutionary corrections to shift the surface brightness (density) $\mu$ profiles to the rest-frame at $z=0$.

{\it Young Objects:} Three systems in this group have composite structure, visible only in the HST images. 
SDSSJ112518.89-014532.4 is a peculiar young object with only a weak $[$O~II$]$ line (section~\ref{sec:spectralfit}) observed with the HST/WFC3 program targeting recently quenched galaxies with high-velocity gas outflows (Proposal ID 12272: C.Tremonti). 

The best-fit model surface brightness profile combines two functions: a compact and an extended S\'ersic profile. The largest half-light radius exceeds the one for the best-fit single S\'ersic model by a factor of nine (R$_{e,c}^{\mathrm{multi}}=6.54$~kpc vs. R$_{e,c}^{\mathrm{single}}=0.74$~kpc, Table~\ref{tab2}). The residual map in Fig.~\ref{fig1} shows spiral structure that we do not try to model. Based on the difference  between the observed and model surface brightness profiles (Fig~\ref{fig1}), at galactocentric distances R$_e\gtrsim~3$~kpc~$\approx4.5\times$R$_{e,c}^{\mathrm{single}}$ the multi-profile model describes the observed light profile at low surface brightness levels considerably better than the single S\'ersic model. The second component makes no significant difference in the core.

HST images of the other two young systems exhibit similarly complex rest-frame optical morphology.  In the most extreme case (SDSSJ150603.69+613148.1) two objects appear in the SDSS images as a single point source. We attribute the SDSS spectrum of this double system, dominated by a young-age SSP, to the larger object containing most of the observed light with the best-fit surface brightness profile parameters in Table~\ref{tab2}. 

SDSSJ112518.89-014532.4  (Fig~\ref{fig1}) has the highest velocity dispersion in this subsample ($\sigma(\mathrm{SDSSJ112518.89-014532.4})> 200$~km~s$^{-1}$, Table~\ref{tab1}) and its  morphology is bulge-dominated. The light profiles for the other two  young galaxies are disk-like.

{\it Old Objects :} These galaxies generally display much smoother surface brightness profiles. 
SDSSJ123657.44+631115.4 resides in one of the X-ray luminous clusters targeted by multi-wavelength HST ACS/WFC3 snapshot survey (Proposal ID 12166: H. Ebeling). The abundant  HST imaging allows  construction of the best-fit 2D models in three filters (F606W, F814W and F110W). 

The light profiles in all three bands have very similar S\'ersic parameters (R$_{e,c}^{\mathrm{multi}}\approx3$~kpc, R$_{e,c}^{\mathrm{single}}\approx1.7$~kpc). We obtain a good fit with only two S\'ersic profiles (Table~\ref{tab2}). The first row of Figure~\ref{fig2} shows an RGB image composed of two HST ACS images and one rescaled HST WFC3 image. We also plot the corresponding best-fit 2D models and residuals. Radial surface brightness profiles of the models in all three wavelength bands closely follow the observed profile out to $3-4$~R$_{e,c}$.  We note that  modeling light profiles for galaxies like this target that are close to the brightest cluster galaxy (BCG) depends critically on the simultaneous modeling of the light profiles for all neighboring systems, including the two-component BCG \citep[e.g., ][]{Gonzalez2005}. 

The HST morphology of the  other five old systems have similar structure.  Most of these systems have round shapes and spheroid-like single S\'ersic profiles with $n_{\textrm{\scriptsize S\'ersic}}^{\mathrm{single}}>2.5$. 

Like their high-$z$ counterparts \citep[e.g.][]{Bruce2012,Buitrago2012}, three of these objects have some disk component. One of them,  SDSSJ001619.07-003358.8,  appears to be an old system with a face-on disk. Based on the best-fit multi-S\'ersic model, the surface brightness profile of this system is a combination of two disk profiles.  The observed profiles of two other galaxies, SDSSJ123130.98+123224.2 and SDSSJ132950.58+285254.8, display  edge-on disk components. The visible light of  SDSSJ123130.98+123224.2 is mostly distributed within a disk-like profile (its bulge-to-total ratio is $B/T<50\%$, based on the best two-component fit). SDSSJ132950.58+285254.8  is a bulge-like object with a weak extended disk ($B/T>60\%$).

\section{Structural and Dynamical Properties of Compact Systems at Intermediate Redshifts}\label{sec:size mass}

We combine the parameters of the best-fit single S\'ersic model for each galaxy (from Table~\ref{tab2}) with  measured velocity dispersions (from Table~\ref{tab1})  to derive dynamical masses:

\begin{equation} 
\mathrm{M}_{dyn}=\frac{\beta(n_{\textrm{\scriptsize S\'ersic}}^{\mathrm{single}})\sigma^2 R_{e,c}^{\mathrm{single}}}{G}, \label{eq:mdyn}
\end{equation}

\noindent where $\beta(n_{\textrm{\scriptsize S\'ersic}}^{\mathrm{single}})$ is a function of the S\'ersic index \citep{Cappellari2006}:

\begin{equation}
\beta(n_{\textrm{\scriptsize S\'ersic}}^{\mathrm{single}})=8.87-0.831n_{\textrm{\scriptsize S\'ersic}}^{\mathrm{single}}+0.024(n_{\textrm{\scriptsize S\'ersic}}^{\mathrm{single}})^2. \label{eq:beta}
\end{equation}

We compare the relations between structural and dynamical properties of our $0.2<z<0.6$ sample with the results obtained in two different redshift regimes: $z\lesssim0.3$ and $0.8<z<2.2$. The structural parameters of single-S\'ersic models for the low-redshift sample \citep[Table~3; ][]{Simard2011} represent 2D decompositions of the $g-$ and $r-$band surface brightness profiles for resolved systems in SDSS~DR7. Again we require  EW$[$O~II$]<5$~\AA \ (see section~\ref{sec:data}).  For the $z\sim0$ sample retrieved from the SDSS~DR7 database we also require\footnote{as recommended at \url{http://www.sdss.org/dr7/products/spectra/}}: i) an average signal-to-noise ratio per pixel of $S/N>10$ and ii) a measured velocity dispersions  in the range $70\, \mathrm{km~s^{-1}}<z<420\, \mathrm{km~s}^{-1}$.   We include small samples of compact $z\sim0$ galaxies described in \citet{Trujillo2009} and \citet{Taylor2010}. The high-redshift comparison sample is a collection of high-resolution HST imaging and spectroscopic data compiled by \citet{VandeSande2013}. All systems in this sample are quiescent galaxies with dynamical masses of M$_{dyn}>2\times10^{10}$~M$_\odot$. We correct measured velocity dispersions for all three samples using the model provided in \citet{VandeSande2013}.  

The left-hand panel of Figure~\ref{fig3} clearly demonstrates the difference in size between galaxies with similar velocity dispersions at $z\sim0$ and at $z>0.2$. For the main $z\sim0$ comparison sample (gray histogram), the median velocity dispersion is 185~km~s$^{-1}$.  For our sample (cyan; $z \sim 0.4$) the median is 178~km~s$^{-1}$ and for $z >1$ (red) the median is 260~km~s$^{-1}$. The corresponding  median sizes are 5.9~kpc, 0.74~kpc, and 2.2~kpc, respectively. The median size of $z\sim0$ sample selected to be compact (gray points) is 1.5~kpc. Although the sizes of our intermediate-redshift compact galaxies are several times smaller than for the $z\sim 0$ systems, velocity dispersions of two samples span the same range of values.

Most of our sample lies in the lower portion of the velocity dispersion range covered by the $z>1$ sample, but they follow the same trend in size-velocity dispersion parameter space. Furthermore, the intermediate-redshift object with the highest velocity dispersion (Fig.~\ref{fig2}) falls very close to the locus of the high-redshift sample.

We note that multiple-profile models tend to overestimate the size of the young morphologically disturbed systems in order to fit their extended asymmetric low-surface-brightness features.   
Thus the upper limits on the half-light radii derived from the best multi-S\'ersic fits can bring three young compact galaxies in our sample very close to the locus of low-redshift galaxies, but they may be misleading (see the 2D profile in Figure \ref{fig1}). 

The right-hand panel of Figure~\ref{fig3} shows a tight size-dynamical mass relation in all three redshift regimes. This relation includes additional information about the S\'ersic index of the best-fit profiles (equations~\ref{eq:mdyn}~and~\ref{eq:beta}). As noted by e.g. \citet{VandeSande2013}, there is a clear offset between the loci of $z\sim0$ and $z>1$ galaxies in the range of dynamical masses where the two samples overlap ($2\times10^{10}$~M$_\odot\leqslant$~M$_{dyn}\leqslant1.86\times10^{12}$~M$_\odot$). In contrast, the sample of compact $z\sim0$ galaxies overlaps with high-redshift systems.  Although our intermediate-redshift galaxies have an average dynamical mass lower than high-redshift quiescent systems ($7.95\times10^{9}$~M$_\odot\leqslant$~M$_{dyn}(z\sim0.4)\leqslant6.31\times10^{11}$~M$_\odot$), the two samples follow the same size-dynamical mass relation. The half-light radius of our most extreme target with $\sigma>300$~km~s$^{-1}$ (Fig.~\ref{fig2}) is very similar to or even smaller than the average size of similarly massive high-redshift systems. This result suggests that M$_{dyn}\approx10^{10}$~M$_\odot$ compact systems should also exist at $z>1$, but with half-light radii of R$_{e,c}\approx0.5$~kpc (or $0\farcs05$ at $z=1$) and with currently undetectable  extended low surface brightness features. 

\section{Conclusions}

We identify nine galaxies with dynamical masses of M$_{dyn}\gtrsim10^{10}$~M$_\odot$  as photometric point sources, but with redshifts between $0.2<z<0.6$  in the SDSS spectro-photometric database. These nine galaxies  have archival HST images demonstrating that they are indeed extremely compact. 

It is imperative to track the change in number density of compact systems with redshift, but no meaningful constraint can be derived from our inhomogeneous, serendipitous sample (see Tables~\ref{tab1}~and~\ref{tab2}). Our sample, however, demonstrates existence: larger samples of 
intermediate redshift compact quiescent galaxies based on well-defined selection criteria should provide number density estimates. 

In size-dynamical-mass parameter space our nine compact galaxies lie away from the typical $z\sim0$ SDSS galaxies of similar mass. The most massive system in our sample  - SDSSJ123657.44+631115.4 -  lies right within the locus of massive compact $z>1$ galaxies. The existence of these
serendipitously discovered intermediate redshift compact galaxies provide clues to uncovering larger samples  for determining the evolution of dense systems routinely observed at high redshift. 

\acknowledgments
We acknowledge the use of the SDSS DR7 data (\url{http://www.sdss.org/dr7/}) and the MAST HST database (\url{http://archive.stsci.edu/hst/}). We thank the referee for prompt, helpful comments. ID is supported by the Harvard College Observatory Menzel Fellowship and NSERC (PDF-421224-2012). The Smithsonian Institution supports the research of IC, HSH, and MJG. IC acknowledges support from grant MD-3288.2012.2.

\clearpage

\begin{deluxetable*}{lccccccc}[h!]
\tabletypesize{\scriptsize}
\tablecaption{Spectroscopic properties\label{tab1}}
\tablewidth{0pt}
\tablehead{
\colhead{ID} & \colhead{$z$} & \colhead{$\sigma_{obs}$} & $\frac{\sigma_{obs}}{\sigma}$ & \colhead{Age} & \colhead{[Z/H]} & \colhead{SDSS Program/Target}  & \colhead{Comment}  \\
\colhead{} & \colhead{} & \colhead{[km~s$^{-1}$]} & & \colhead{[Myr]} & \colhead{[dex]} & \colhead{} &  \colhead{} \\
\colhead{(1)} & \colhead{(2)} & \colhead{(3)} & \colhead{(4)} & \colhead{(5)} & \colhead{(6)} & \colhead{(7)} & \colhead{(8)}
} 
\startdata
SDSSJ001619.07-003358.8  &   0.37199   &   $114\pm22$  & 0.9268/0.9333 &    $538\pm18$ & $0.56\pm0.15$  & Southern/ & Old (E+A)\\
&&&&&&QSO\_HIZ&\\
SDSSJ100218.66+143757.0  &   0.33049   &   $272\pm14$  & 1.0015/1.0015 &  $>8900$ & $-0.05\pm0.07$  & Legacy/ & Old  \\
&&&&&&GALAXY\_RED&\\
SDSSJ112518.89-014532.4  &   0.51926   &   $210\pm14$  &  0.9225/0.9880 &  $35\pm2$ & $-0.53\pm0.11$ & Legacy/ & Young \\  
&&&&&&QSO\_SKIRT&\\
SDSSJ123130.98+123224.2  &   0.41553   &   $178\pm19$  &  0.9207/0.9348 &  $1448\pm122$ & $-0.33\pm0.19$ & Legacy/ & Old \\
&&&&&&QSO\_HIZ&\\
SDSSJ123657.44+631115.4  &   0.29884   &   $331\pm22$  & 0.9480/0.9650 &   $>8900$ & $-0.01\pm0.08$  & Legacy/ & Old\\
&&&&&&GALAXY\_RED, &\\
&&&&&&ROSAT\_D&\\
SDSSJ132950.58+285254.8  &   0.35661   &   $197\pm27$  &  0.9285/0.9346 &  $1229\pm161$ & $0.22\pm0.20$  & Legacy/ & Old \\
&&&&&&QSO\_HIZ&\\
SDSSJ135920.98+513738.9  &   0.41292   &   $159\pm24$  &  0.9115/0.9509 &  $20\pm1$ & $0.14\pm0.13$  & Legacy/ & Young \\
&&&&&&QSO\_CAP, &\\
&&&&&&SERENDIP\_BLUE&\\
SDSSJ143026.34+342944.6  &   0.44704   &   $135\pm26$  &  0.9142/0.9142 &   $954\pm85$ & $-0.06\pm0.23$ & Legacy/ & Old (E+A) \\
&&&&&&QSO\_HIZ&\\
SDSSJ150603.69+613148.1  &   0.43680   &   $145\pm24$  & 0.9086/0.9336 &   $32\pm1$ & $-0.20\pm0.15$  & Legacy/& Young \\
&&&&&&QSO\_SKIRT&\\
\enddata
\tablecomments{Columns: (1) SDSS designation; (2) Redshift; (3) Observed velocity dispersion; (4) Aperture correction for single-/multi-S\'ersic profile (see~Table~\ref{tab2}); (5) Age of the best-fit SSP model; (6) Metallicity of the best-fit SSP model; (7) SDSS target flag; (8)  Classification}
\end{deluxetable*}

\begin{deluxetable*}{lccccccccc}[h!]
\tabletypesize{\scriptsize}
\tablecaption{Structural properties\label{tab2}}
\tablewidth{0pt}
\setlength{\tabcolsep}{0.02in} 
\tablehead{
\colhead{ID} & \colhead{R$_{e,c}^{\mathrm{single}}$} & \colhead{$n_{\textrm{\tiny S\'ersic}}^{\mathrm{single}}$} & \colhead{$(b/a)^{\mathrm{single}}$} & \colhead{R$_{e,c}^{\mathrm{multi}}$} & \colhead{$n_{\textrm{\tiny S\'ersic}}^{\mathrm{multi}}$} & \colhead{$(b/a)^{\mathrm{multi}}$} & \colhead{$N$} & \colhead{HST program/camera/filter} & \colhead{Comment}  \\
\colhead{} & \colhead{[kpc]} & \colhead{} & \colhead{} & \colhead{[kpc]} & \colhead{} & \colhead{} &  \colhead{} & \colhead{(pixel scale [$\arcsec$/pix])} & \colhead{} \\
\colhead{(1)} & \colhead{(2)} & \colhead{(3)} & \colhead{(4)} & \colhead{(5)} & \colhead{(6)} & \colhead{(7)} & \colhead{(8)} & \colhead{(9)} & \colhead{(10)}
}
\startdata
SDSSJ0016-0033  & 0.74 &     1.63 &      0.78  &      1.07 &    0.91 &   0.92 &   2 &   10555/ACS/F625W (0.05) & disk \\
SDSSJ1002+1437  & 6.53 &     4.58 &      0.85 &      6.53 &    4.58 &   0.85 &   1 &   10084/WFPC2/F606W (0.10) & bulge \\
SDSSJ1125-0145  & 0.74 &     8.00 &      0.90 &      6.54 &     4.26 &   0.70 &   2 &   12272/WFC3 /F814W (0.04) & bulge-dominated \\
SDSSJ1231+1232  & 0.61 &     3.42 &      0.16 &      1.24 &    1.04 &  0.14 &   2 &   5370/WFPC2/F814W (0.10) & disk-dominated \\
SDSSJ1236+6311~or   & 1.76 &     2.53 &      0.94 &      3.25 &    2.76 &  0.95 &    2 &   12166/ACS/F606W (0.05) & bulge-dominated \\
SDSSJ1236+6311~or  & 1.75 &     2.53 &      0.94 &      2.81 &    2.03 &    0.98 & 2 &   12166/ACS/F814W (0.05) & bulge-dominated \\ 
SDSSJ1236+6311  & 1.65 &     2.20 &      0.94 &      2.55 &    2.78 &     0.98 & 2 &   12166/WFC3/F110W (0.128) & bulge-dominated \\
SDSSJ1329+2852  & 0.79 &     2.65 &      0.38 &      1.13 &    0.25 &      0.16 & 3 &   10626/ACS/F814W (0.05) & bulge-dominated \\ 
SDSSJ1359+5137  & 0.45 &     5.47 &      0.86 &      2.36 &    4.09 &      0.89 & 4 &   12019/WFC3/F814W (0.04) & disk-dominated\tablenotemark{a}  \\ 
SDSSJ1430+3429  & 0.51 &     3.36 &      0.53 &      0.51 &    3.36  &   0.53 &   2 &   10890/WFPC2/F606W (0.1) & bulge-dominated\tablenotemark{b} \\ 
SDSSJ1506+6131  & 0.43 &     8.00 &      0.75 &       1.21 &    2.30 &   0.64 &  2 &   12019/WFC3/F814W (0.04) & disk-dominated \\
\enddata
\tablecomments{Columns: (1) SDSS designation (abridged); (2) Circularized half-light radius of the single-profile model (R$_{e,c}^{\mathrm{single}}=\mathrm{R}_e^{\mathrm{single}}\times \sqrt{b/a}$, where R$_e$ is the major axis half-light radius and $b/a$ is the axial ratio); (3) S\'ersic index of the single-profile model; (4) Axial ratio of the single-profile model;  (5) The largest (circularized) half-light radius of the composite model; (6) S\'ersic index corresponding to (5); (7) Axial ratio corresponding to (5); (8) Number of profiles in the composite model; (9) HST program ID, camera and filter (corresponding pixel scale); (10) Morphology.}
\tablenotetext{a}{The largest component in this multi-S\'ersic fit has a de Vaucouleurs profile, but the inner three disk-like components with $n\approx2$ dominate the observed light profile.}
\tablenotetext{b}{ This object requires two S\'ersic profiles fitted simultaneously:  a disk and a more prominent bugle with similar half-light radii.}
\end{deluxetable*}

 \end{document}